\documentclass[aps,showpacs,twocolumn,pra]{revtex4-1}

\usepackage{amsmath}
\usepackage{amssymb}
\usepackage{graphicx}
\usepackage{bm}
\usepackage{color}

\begin{document}
\title{Locality of the Aharonov-Bohm-Casher effect}


\author{Kicheon Kang}
\email{kicheon.kang@gmail.com}
\affiliation{Department of Physics, Chonnam National University,
 Gwangju 500-757, Republic of Korea}

\begin{abstract}
We address the question of the locality versus nonlocality
in the Aharonov-Bohm and the Aharonov-Casher effects.
For this purpose, we investigate all possible configurations
of ideal shielding of the overlap between the electromagnetic fields 
generated by a charge and by a magnetic flux, and analyze their
consequences on the Aharonov-Bohm-Casher interference.
In a classical treatment of shielding, the Aharonov-Bohm-Casher effect 
vanishes regardless of the geometry of shielding, 
when the local overlap of electromagnetic fields is completely eliminated. 
On the other hand, the result depends on the configuration of shielding 
if the charge quantization in the superconducting shield is taken
into account.
It is shown that our results are fully understood in terms of 
the fluctuating local-field interaction. 
Our analysis strongly supports
the alternative view on the Aharonov-Bohm-Casher interference that the
effects originate from the local action of electromagnetic fields. 
\end{abstract}
\maketitle

\section{introduction}
It is a widely accepted view that the Aharonov-Bohm~(AB) effect~\cite{ab59}, 
which describes the quantum interference of a charged particle moving 
in a region free of electromagnetic fields, 
is a pure topological phenomenon that cannot be understood 
from the local action of fields.
In contrast to this common viewpoint, we have recently shown that 
a fully local description of the force-free AB effect is possible 
based on the Lorentz-invariant interactions of electromagnetic 
fields~\cite{kang13}. 
Naturally, this resolves a long-standing puzzle on why 
the old local-interaction-based theories~\cite{liebowitz65,boyer73,spavieri92} 
have failed to provide a consistent description of the force-free AB effect:
The relativity principle was not
properly taken into account in the previous field-interaction-based theories,
and neglecting the Lorentz invariance gives rise to erroneous classical
forces.
An immediate and important corollary of the local-interaction-based framework 
is that the AB
effect vanishes if all local overlap of electromagnetic fields is 
shielded.
The purpose of this paper is to provide an extensive investigation on 
the effect of shielding
and its consequences on the AB effect and the Aharonov-Casher(AC) effect~\cite{ac84} - a dual
phenomenon to the AB effect.

The main question raised concerning the standard nonlocal viewpoint can be
clearly stated as follows, which was already noticed in 
Ref.~\onlinecite{reznik89}. Let us consider a charge ($q$) and 
a fluxon~($\Phi$) in 2+1 dimension~(Fig.~1). 
The principle of relativity demands that the physical law governing the system
be independent of the reference frame. The AB effect is described in the
frame of the stationary fluxon, $O_\Phi$~(Fig.1(a)). 
The AB phase shift is induced by the charge
encircling the fluxon. The standard viewpoint on this phenomenon is that
it is purely topological and originates from the nonlocal interaction of
the charge and the fluxon, described by the ``nonlocal" Lagrangian
\begin{equation}
 L_i^{nl} = \frac{q}{c}\dot{\mathbf{x}}\cdot \mathbf{A}(\mathbf{x}) ,
\label{eq:Li_nl}
\end{equation}
where $\mathbf{x}$ and $\mathbf{A}(\mathbf{x})$
are the position of the charge relative to the fluxon
and the vector potential generated by the fluxon, respectively.
The ``nonlocality" here implies that the charge and the magnetic field
do not experience a direct local interaction.
In this framework, the AB effect is purely topological in that it requires
that the charge be confined in a multiply connected region and in that
there is no realistic way to relate a phase shift to any particular
position in the charge's path.

On the other hand, in the stationary charge's reference frame, 
$O_q$~(Fig.1(b)),
the moving fluxon acquires
a phase shift via the local interaction Lagrangian
\begin{equation}
 L_i^l = \frac{\dot{\bar{\mathbf{x}}}}{4\pi c}\cdot
       (\Phi\hat{z}) \times \mathbf{E}(\bar{\mathbf{x}}) ,
\label{eq:Li_l}
\end{equation}
where $\hat{z}$ is the unit vector perpendicular to the plane, and
 $\bar{\mathbf{x}}=-\mathbf{x}$ is the position of the fluxon relative to
the charge. $\mathbf{E}(\bar{\mathbf{x}})$
is the electric field generated by the charge at the location of the fluxon.
The phase shift that appears in this reference frame is known as 
the AC effect~\cite{ac84}. It can be understood in terms of 
the local-interaction-induced accumulation of the phase shift,
although it is rarely noticed.
The locality of the AC effect was also demonstrated 
in Ref.~\onlinecite{peshkin95}.

In two spatial dimension, however, the AB and the AC effects are actually 
the same phenomena - only their reference frames are different. 
The duality of the two phenomena is demonstrated by the equality 
of the phase accumulated by one loop rotation derived in each frame:
\begin{equation}
 \varphi = \frac{q}{\hbar c} \oint \mathbf{A}(\mathbf{x})\cdot d\mathbf{x}
   = \frac{\Phi}{4\pi\hbar c} \oint \hat{z} \times 
        \mathbf{E}(\bar{\mathbf{x}})\cdot d\bar{\mathbf{x}}
   = \frac{q\Phi}{\hbar c} \,.
\end{equation}
In spite of this equality, the result would be interpreted differently
for the two observers in the frames $O_\Phi$
and $O_q$, respectively. Intriguingly, third
observer ($O'$), who finds that both particles are moving~(Fig.~1(c)), 
would be frustrated with
the two contradictory interpretations. 
This observer cannot even decide with which
Lagrangian, Eq.(\ref{eq:Li_nl}) or Eq.(\ref{eq:Li_l}), to start.
As noticed in Ref.~\cite{reznik89}, 
there would be two possible resolutions to this inconsistency: 
Either (i) the AC effect should be interpreted as a nonlocal and purely 
topological effect or 
(ii) the AB effect is a result of local interaction. 
The common attitude to this problem is to ignore the paradox, or, to regard 
possibility (i) as the solution, which has also been argued in
Ref.~\onlinecite{reznik89} for a particular arrangement of the AC setup.

In this paper, in contrast to the common notion, we show that 
resolution (ii) is possible and that it provides a more universal 
framework independent of the reference frame.  
Introducing various types of shielding the overlap of electromagnetic fields, 
we show that the AB effect vanishes if the local interaction of the fields 
is completely eliminated. Our results strongly support the validity of the 
alternative local-interaction-based theory of 
the Aharonov-Bohm-Casher~(ABC) effect.

The paper is organized as follows. First, the field-interaction-based 
Lagrangian and Hamiltonian are introduced for a charge and a fluxon in 
two dimension (Section II). 
In Section III, a classical treatment of ideal shielding is briefly discussed,
and it is shown that the ABC interference disappears if the local overlap of the
fields produced by the two particles vanishes.
Then, we provide an extensive quantum mechanical treatment in the
presence of superconducting
barriers placed in between the charge and the fluxon (Section IV). The 
result depends on the specific configuration of the system, whereas 
we find that
the ABC effect is determined by the fluctuating local field interaction 
in general.
The implications of the results in Sections III and IV are summarized and 
discussed in Section V. Our conclusion is given in Section VI. 

\section{Field-interaction Lagrangian for a charge and a fluxon}
The simplest case that can be conceived for describing 
the ABC effect is a two-particle system of 
a charge $q$ and a fluxon $\Phi$ at locations of $\mathbf{r}$ and
$\mathbf{R}$, respectively.
Our starting point is to introduce the universal Lorentz-covariant field 
interaction Lagrangian for
describing the interaction between the two particles~\cite{kang13}. 
An obvious merit of this approach is that we do not have to worry about 
choosing which of the two pictures, nonlocal~(Eq.~(\ref{eq:Li_nl})) 
or local~(Eq.~(\ref{eq:Li_l})), as our starting point. 
In the limit of $\dot{r},\dot{R}\ll c$, 
the Lagrangian of the system is given by~\cite{kang13}
\begin{subequations}
\label{eq:local-lagrangian}
\begin{equation}
  L = \frac{1}{2} m \dot{\mathbf{r}}\cdot\dot{\mathbf{r}}
   + \frac{1}{2} M \dot{\mathbf{R}}\cdot\dot{\mathbf{R}}
   + L_{\rm int},
\end{equation}
where $m$($M$) is the mass of the charge(fluxon). 
The interaction between the two particles, $L_{\rm int}$,
is derived from the Lorentz-covariant field interactions~\cite{kang13} as
\begin{equation}
 L_{\rm int} = \frac{1}{4\pi} \int
     \left( \mathbf{B}_q\cdot\mathbf{B}_\Phi - \mathbf{E}_q\cdot\mathbf{E}_\Phi
     \right) \,d\tau \,,
\label{eq:Lint}
\end{equation}
where $\mathbf{B}_q$($\mathbf{E}_q$) and $\mathbf{B}_\Phi$($\mathbf{E}_\Phi$)
are the magnetic(electric) fields produced by the charge and by the fluxon, 
respectively. Note that this Lagrangian is uniquely constructed 
from the obvious first principles of 
(i) relativity, (ii) linearity in the field strengths, 
and (iii) correspondence with the known result in the limit of stationary 
charge~\cite{kang13}. 
One can rewrite Eq.~(\ref{eq:Lint}) in the simpler form
\begin{equation}
 L_{\rm int} = (\dot{\mathbf{r}} - \dot{\mathbf{R}})\cdot \mathbf{\Pi} \,,
\label{eq:LintPi}
\end{equation}
with the field momentum $\mathbf{\Pi}$ produced 
by the overlap of charge's electric field 
and the fluxon's magnetic field:
\begin{equation}
\mathbf{\Pi} = \frac{1}{4\pi c} \int \mathbf{E}_q\times \mathbf{B}_\Phi d\tau \,.
\end{equation}
Note that Eq.~(\ref{eq:LintPi}) is equivalent to the interaction Lagrangian
\begin{equation}
 L_{\rm int} = \frac{q}{c} (\dot{\mathbf{r}} - \dot{\mathbf{R}})\cdot \mathbf{A} \,,
\end{equation}
based on the vector potential $\mathbf{A}$, except that in the former case
(Eq.~(\ref{eq:LintPi})), the Lagrangian is given only by the gauge-independent
physical quantities.
\end{subequations}

This Lagrangian (Eq.~\ref{eq:local-lagrangian}) can also be transformed to 
the Hamiltonian 
\begin{equation}
 H = \frac{1}{2m} \left( \mathbf{p}-\mathbf{\Pi} \right)^2
   + \frac{1}{2M} \left( \mathbf{P}+\mathbf{\Pi} \right)^2 \,,
\label{eq:local-hamiltonian}
\end{equation}
where $\mathbf{p}$ and $\mathbf{P}$ are the canonical momenta of the charge
and the fluxon, respectively.
A noticeable point about these canonical momenta is that they are physical
quantities given by
\begin{equation}
 \mathbf{p} = m\dot{\mathbf{r}}+\mathbf{\Pi},\;\;\;
  \mathbf{P} = M\dot{\mathbf{R}}-\mathbf{\Pi},
\end{equation}
contrary to their gauge dependence in the standard potential-based model.
The canonical momentum of charge, $\mathbf{p}$, 
is the sum of the mechanical ($m\dot{\mathbf{r}}$)
and the field ($\mathbf{\Pi}$) momenta. 
The canonical momentum of the fluxon, $\mathbf{P}$, can
be interpreted as the sum of the translational ($M\dot{\mathbf{R}}$) 
and the hidden relativistic~\cite{aharonov88} mechanical momenta. 
In both cases, the canonical momenta are 
the net momenta carried by each particle.

It has previously been shown~\cite{kang13} that
this local-interaction-based 
Lagrangian~(Hamiltonian) of Eq.~(\ref{eq:local-lagrangian})
(Eq.~(\ref{eq:local-hamiltonian})) reproduces 
the main features of the ABC effect;
(i) the absence of the classical mechanical force and (ii) the 
appearance of the
topological phase $\varphi$ 
(formed by the local accumulation of the field momentum in our 
framework),
\begin{equation}
 \varphi = \frac{1}{\hbar} \oint\mathbf{\Pi}\cdot d\mathbf{r}  
      = q\Phi/\hbar c \,.
\end{equation}

\section{Classical treatment of shielding}
To answer the question of locality versus nonlocality of the ABC effect, 
it is essential to investigate
the case where local overlap of the electromagnetic fields is eliminated. 
For this purpose, we consider three
possible configurations of shielding~(Fig.~2). In all cases, an ideal 
Faraday cage is placed between the charge and the fluxon. 
We adopt the local-interaction-based Lagrangian and Hamiltonian of 
Eqs.~(\ref{eq:local-lagrangian}) and (\ref{eq:local-hamiltonian}) 
for this analysis,
but it should be noted that the standard potential-based description 
leads to the same result. Note that the shielding configuration realized
in the experiment of Ref.~\onlinecite{tonomura86} cannot be applied to
our analysis here because of the nonadiabatic nature of the field generated
by the incident electrons with high kinetic energy~\cite{kang13}.
(This point will be discussed in Section V.)

Let us first analyze the induced ABC phase in the presence of
an ideal classical conductor placed between the two particles. 
Quantum treatments for
superconducting shields will be provided in the next section. 
In any configuration of the two particles and a conducting shield~(Fig.~2), 
the interactions are governed by the overlaps 
among the electromagnetic fields produced by
the charge, the fluxon, and the conducting shield. 
The only constraint imposed here is that
the induced charge density in the conducting shield generates 
an electric field ($\mathbf{E}_s$) that
compensates for the field of the charge~\cite{floating-sc}. 
Then, we find the net interaction Lagrangian as
\begin{equation}
 L_{\rm int} = \frac{1}{4\pi} \int
   \left[  (\mathbf{B}_q+\mathbf{B}_s)\cdot\mathbf{B}_\Phi 
         - (\mathbf{E}_q+\mathbf{E}_s)\cdot\mathbf{E}_\Phi
   \right] \,d\tau + L_{qs}\,,
\label{eq:Lint-shield}
\end{equation}
where $\mathbf{B}_s$($\mathbf{E}_s$) is the magnetic(electric) field
generated by the induced charge in the superconductor.
$L_{qs}$, which represents the interaction between the charge $q$ and 
the conducting shield, is irrelevant to the ABC effect. 
For an ideal classical conductor, the electric and magnetic fields 
generated by charge $q$ are exactly canceled by the induced
charges in the conducting shield, that is, 
$\mathbf{E}_q + \mathbf{E}_s=0$ and $\mathbf{B}_q + \mathbf{B}_s=0$,
at the location of the fluxon. 
Therefore, the interaction
Lagrangian of Eq.~(\ref{eq:Lint-shield}) is independent of the localized 
magnetic flux, and
accordingly the ABC effect vanishes. Note that this conclusion is valid
also in the potential-based treatment~\cite{kang13}, which has been
widely overlooked.

\section{Quantum treatment of shielding}
In the previous section, we have shown that the ABC phase shift vanishes 
in the classical treatment
of ideal shielding, independent of the geometry considered in Fig.~2.
Whereas this already demonstrates the locality of the ABC effect, 
it is important to test the validity of the locality with a quantum
mechanical treatment of the conducting shield.
In the following, we show that the effect of shielding depends on 
the geometry of the system, when we take
into account the quantization of charges in the superconductor.
For a quantum treatment, we consider an ideal superconductor placed in
between the charge and the fluxon. Ideal shielding in the quantum
treatment implies that
the quantum mechanical average values of the net electric and 
magnetic fields,  
generated by charge $q$ and the induced charges in the superconductor,
vanish at the location of the fluxon.

\subsection{Configuration I: Fluxon confined in a superconductor}
In Configuration I~(Fig.~2(a)), 
the fluxon is confined inside the superconducting
Faraday cage, and the charge is moving in the region outside the 
superconductor. Here we use the potential-based theory for convenience,
but as we have already pointed out 
in the previous section, 
the same conclusion is drawn from
the field-interaction-based framework. 
The Lagrangian of the system is given by
\begin{subequations}
\begin{equation}
 L = \frac{m}{2} \dot{\mathbf{r}}\cdot\dot{\mathbf{r}} + L_{\rm sc} +
     L_{\rm int} \,,
\end{equation}
where $L_{\rm sc}$ describes the superconductor, and the interaction
term 
\begin{equation}
 L_{\rm int} = \frac{q}{c}\dot{\mathbf{r}}\cdot\mathbf{A} 
   + \frac{1}{c}\int \delta n(\phi) \,\mathbf{v}_c\cdot\mathbf{A}\, 
     d\mathbf{r}' ,
\label{eq:Lint1}
\end{equation} 
\label{eq:lagrangian1}
\end{subequations}
is composed of two parts: The first term is the main interaction between
the charge $q$ and the fluxon; the second term represents the interaction
between the superconductor and the fluxon. 
The moving charge $q$ induces a variation of the surface charge density,
$\delta n(\phi)$, on the outer surface of
the superconductor moving with velocity $\mathbf{v}_c$. 
$\delta n(\phi)$ depends on the azimuthal angle $\phi$ 
of the position vector $\mathbf{r}'$ on the surface (see Fig.~3). 
The interaction between charge $q$ and the superconductor, which is
irrelevant to the ABC phase shift, is neglected here.

The transition amplitude of the moving charge $q$ 
from a point $(\mathbf{r}_i,t_i)$ to another $(\mathbf{r}_f,t_f)$
in spacetime is given by
\begin{eqnarray}
 u_{i\rightarrow f} 
   &=& \langle\mathbf{r}_f|\otimes\langle\eta_f| e^{-iH(t_f-t_i)/\hbar}
       |\mathbf{r}_i\rangle\otimes|\eta_i\rangle   
   \label{eq:amplitude} \\
   &=& \lim_{M\rightarrow\infty} 
       \langle\mathbf{r}_f|\otimes\langle\eta_f| 
         \left(e^{-iH\epsilon/\hbar} \right)^M
       |\mathbf{r}_i\rangle\otimes|\eta_i\rangle ,
   \nonumber
\end{eqnarray} 
where $|\eta_i\rangle$ ($|\eta_f\rangle$) denotes the initial (final) 
state of the
superconducting shield, and $\epsilon \equiv (t_f-t_i)/M$.
The state of the superconducting shield can be written as
\begin{equation}
 |\eta\rangle = \sum_m b_m|\psi_m\rangle \,,
\label{eq:eta}
\end{equation}
where $|\psi_m\rangle$ stands for a substate with $m$ excess Cooper pairs on the outer surface of the
superconductor. The state $|\psi_m\rangle$ satisfies
\begin{equation}
 \int d\mathbf{r}' \langle\psi_m|n(\phi)|\psi_m\rangle = 2em ,
\end{equation}
that is, the induced charge is quantized in units of 2e. 

The transition amplitude of Eq.~(\ref{eq:amplitude}) should
be evaluated by taking into account all possible trajectories of the
charge $q$ and the superconductor. 
In our case, we
are interested only in the paths in which, inside the cage, 
the electric and the magnetic fields
generated by the charge $q$ are perfectly shielded by the
induced charges in the superconductor. 
The question here is whether this condition is fulfilled 
in the presence of quantization of the induced charges (by $2em$). 
One can find that this is indeed the case, 
and the electric and the magnetic fields inside the cage vanish
for any number $m$ of excess Cooper pairs, if and only if 
\begin{subequations}
\begin{equation}
 \langle\psi_m|n(\phi)|\psi_m\rangle = n_0^m + \delta n(\phi) ,
\end{equation}
where $n_0^m = (2em+q)/2\pi R$ ($R$ being the outer radius of the
superconductor) is a constant, 
and the inhomogeneous part,
\begin{equation}
 \delta n(\phi) = -\frac{q}{2\pi R} \left( 
     \frac{r^2-R^2}{r^2+R^2-2rR\cos{\phi}} \right) 
\end{equation}
\label{eq:induced-charge}
\end{subequations}
provides perfect shielding of the fields. 
The transition amplitude of Eq.~(\ref{eq:amplitude}), 
with the above condition (Eq.~(\ref{eq:induced-charge})) 
for the superconducting barrier, can be rewritten 
in the form
\begin{equation}
 u_{i\rightarrow f} = \int_{\mathbf{r}_i}^{\mathbf{r}_f} 
   {\cal D} [\mathbf{r}(t)] 
   \exp{ \left\{ \frac{i}{\hbar}\int L(\mathbf{r},\dot{\mathbf{r}}, t)
         \right\} } \,,
\end{equation}
where the Lagrangian is given by Eq.~(\ref{eq:lagrangian1}).
For this path integration, we are summing all contributions 
satisfying the
condition of Eq.~(\ref{eq:induced-charge}). In this case, one can find that 
the first and the second terms of Eq.~(\ref{eq:Lint1}) are given by
$q\Phi\dot{\phi}/(2\pi c)$ and $-q\Phi\dot{\phi}/(2\pi c)$, respectively,
which cancel each other.
Therefore, $L_{\rm int}=0$, and the fluxon does not contribute to the phase
factor of $u_{i\rightarrow f}$. The ABC effect vanishes completely
in this configuration.

\subsection{Configuration II: Charge confined in a superconducting cage}
Now we turn our attention to Configuration II~(Fig.~2(b)),
where the charge is confined inside the cage 
and the fluxon is located outside. 
We  consider only a grounded superconductor~\cite{floating-sc} where
the field produced by the charge cannot penetrate outside the shield.
As we will show in the following, quantum fluctuation of charge is essential
in this configuration (and also in Configuration III).

For convenience, we adopt the stationary charge's frame ($O_q$). 
The transition amplitude of a moving fluxon is given in the same way
as in Eq.~(\ref{eq:amplitude}), 
\begin{equation}
 u_{i\rightarrow f} 
   = \lim_{M\rightarrow\infty} 
       \langle\mathbf{R}_f|\otimes\langle\eta_f| 
         \left(e^{-iH\epsilon/\hbar} \right)^M
       |\mathbf{R}_i\rangle\otimes|\eta_i\rangle ,
\end{equation} 
where the Hamiltonian is given by 
\begin{equation}
 H = \frac{1}{2M} (\mathbf{P}+\mathbf{\Pi})^2 + H_0 \,.
\end{equation}
Here $H_0$ denotes the contribution from the superconducting condensate and the
interaction between charge $q$ and the superconductor. $H_0$ does not
contribute to the ABC phase and so is ignored here.
The field momentum $\mathbf{\Pi}$ is produced by the overlap of the
net electric field of
the charges 
and the magnetic field of the fluxon ($\mathbf{B}_\Phi$) as
\begin{equation}
 \mathbf{\Pi} = \frac{1}{4\pi c} 
  \int (\mathbf{E}_q+\mathbf{E}_s)\times \mathbf{B}_\Phi d\tau 
 \,,
\end{equation}
where $\mathbf{E}_q$ and $\mathbf{E}_s$ denote the electric fields generated
by charge $q$ and by the superconductor, respectively.
The expectation value of $\mathbf{E}_s$ depends on the state of the
superconducting shield, $|\eta\rangle$, given as a coherent superposition
of the number eigenstates $|\psi_m\rangle$ as in Eq.~(\ref{eq:eta}),
where $m$ denotes the excess number of induced Cooper pairs 
on the surface of the
superconductor. Ideal shielding of the fields
in quantum treatment is imposed by the condition
\begin{equation}
 \langle\eta| \mathbf{E}_q+\mathbf{E}_s |\eta\rangle = 0 \,.
\label{eq:shield}
\end{equation}
An intriguing point here is that,
although the expectation value of the
field vanishes for the ideal superconductor, this perfect shielding 
cannot be achieved 
for any particular substate with definite number $m$, $|\psi_m\rangle$,
unless $q$ is an integer multiple of $2e$. That is, 
$\langle\psi_m| \mathbf{E}_q+\mathbf{E}_s |\psi_m\rangle \ne 0$, in general.
This is the major difference from
Configuration I where the field interaction vanishes for any number $m$ of 
excess Cooper pairs. Therefore, in contrast to Configuration I,
the charge-fluxon interaction is not completely suppressed, 
and the transition amplitude is reduced to 
\begin{subequations}
\begin{equation}
 u_{i\rightarrow f} 
   = \sum_{ml}b_m b_{l}^* u_{lm} (f;i) \,,
\end{equation}
where
\begin{eqnarray}
 u_{lm}(f;i)
  &=& \langle\mathbf{R}_f|\otimes\langle\psi_{l}| 
                    e^{-\frac{i}{\hbar}H(t_f-t_i)}
                  |\mathbf{R}_i\rangle\otimes|\psi_m\rangle \nonumber \\
  &=& \int_{\mathbf{R}_i}^{\mathbf{R}_f} {\cal D}[\mathbf{R}(t)]
      e^ { \frac{i}{\hbar} \int L_{lm}(\mathbf{R},\dot{\mathbf{R}})dt } \,.  
\end{eqnarray}
$L_{lm}$, the element of the moving fluxon's Lagrangian, 
is found to be
\begin{equation}
 L_{lm}(\mathbf{R},\dot{\mathbf{R}}) 
    = \left( \frac{M}{2} \dot{R}^2 - \dot{\mathbf{R}}\cdot\mathbf{\Pi}_m 
      \right) \delta_{lm} \,,
\label{eq:L_lm}
\end{equation}
where $\mathbf{\Pi}_m=\langle\psi_m|\mathbf{\Pi}|\psi_m\rangle$ is the field 
momentum for the particular state $|\psi_m\rangle$.
For a fluxon with a negligible size, we find
\begin{equation}
 \mathbf{\Pi}_m = -\frac{q+2em}{2\pi cR}\Phi\hat{\phi} ,
\end{equation} 
\end{subequations}
and the phase factor acquired for one-loop rotation induced by the interaction
term is 
\begin{subequations}
\begin{equation}
 u_1 = \sum_m|b_m|^2 e^{i\varphi_m}, 
\label{eq:u1}
\end{equation}
where the phase shift $\varphi_m$ for a particular substate $|\psi_m\rangle$
is induced by the local field interaction as
\begin{equation}
 \varphi_m = -\frac{1}{\hbar} \oint \mathbf{\Pi}_m\cdot d\mathbf{R}
   = \frac{(q+2em)\Phi}{\hbar c} \,.  
\end{equation}
Unlike Configuration I, the ABC effect is not completely shielded but modified 
by the superconductor. The change of the phase factor
is manifested by the fluctuating local field interaction represented by the term
$-\dot{\mathbf{R}}\cdot\mathbf{\Pi}_m$ in Eq.~(\ref{eq:L_lm}). 
The phase factor $u_1$ in Eq.~(\ref{eq:u1}) depends on the coefficients
$\{|b_m|\}$, with a constraint imposed by the condition
of complete shielding (Eq.~(\ref{eq:shield})), 
\begin{equation}
 \sum_m m|b_m|^2 = -q/2e \,.
\end{equation}  
\end{subequations}

\subsection{Configuration III: Both particles confined in a superconductor}
A particularly interesting limit in the result of the previous section is
found when $\Phi$ is quantized in units of superconducting
flux quantum, $\Phi=(hc/2e)\times\rm{integer}$. 
In fact, this quantization is
achieved for Configuration III. In this case, the phase factor
of Eq.~(\ref{eq:u1}) is reduced to 
\begin{equation}
 u_1 = e^{iq\Phi/\hbar c} .
\end{equation}
That is, the ABC effect is unaffected by the presence of the superconducting
shield, in spite of complete shielding of the (expectation value of) the 
electromagnetic fields. This case was analyzed in Ref.~\onlinecite{reznik89}. 
Based on this result, the authors in Ref.~\onlinecite{reznik89} argued that 
the AC effect also represents a nonlocal interaction, although it appears
as a local interaction of Eq.~(\ref{eq:Li_l}), and concluded that the
nonlocal picture should be applied in both the AB and the AC effects.
It is clear from our study that this is an incomplete argument. 
Configuration III is only a special case
of shielding the local field interactions, and one cannot draw a general conclusion
that the ABC effect is understood only in terms of the nonlocal picture.
Our analysis of the three possible
different configurations shows that the effect of shielding is not very
universal but depends on the type of shielding, and that the result
is determined by the fluctuating local field interactions. 
 
\section{Discussion}
The results of our analysis on shielding the local interaction of fields are
summarized as follows. First, there is a qualitative difference between
classical
and quantum treatments. In the classical approach to shielding, the local 
field interactions are completely suppressed as far as the shielding of the
field is ideal, and this eliminates the ABC effect. In the quantum treatment
of shielding where the charge quantization is taken into account in 
the superconducting shield, two different types appear depending on the
geometry of shielding. 
The shielding of the electromagnetic field eliminates the ABC effect 
in Configuration I~(Fig.~2(a)), 
whereas the shielding does not completely suppress the effect 
in Configurations II and III~(Figs.~2(b) and (c)). 
The main difference between these two classes is the role
of the fluctuating local field interactions. 
In Configurations II and III, 
although the expectation value of the field generated by charge $q$ 
is compensated for by that of the induced charge
in the superconductor, 
the field interactions are not completely suppressed.
The ABC effect is modified (in Configuration II) or even unaffected 
(in Configuration III) 
in spite of an ideal shielding of the charge's field in the position 
of the fluxon. 
We can conclude that this is due to the fluctuating local field
interactions, and that it demonstrates the locality of the electromagnetic 
interaction in Aharonov-Bohm-Casher interferometry. 

Experimentally, no experiments so far have been performed under the condition
of perfect shielding of the field interactions. The most ideal one was the
experiment performed by Tonomura {\em et al.}~\cite{tonomura86}, where the magnetic
flux is shielded by a superconductor from the moving electron's path. 
Their setup is
basically equivalent to Configuration I where the flux is confined in a
superconducting shield. Contrary to the analysis for Configuration I, 
a clear AB phase shift was
observed despite the presence of the superconducting shield. In this experiment,
however, incident electrons with a speed of about $2.4\times 10^8\rm{m/s}$ 
were used. In fact, no superconducting material can shield the magnetic
field produced by such fast electrons~\cite{kang13}, and the ideal
shielding analysis in Section~IV-A cannot be applied to the experiment in
Ref.~\onlinecite{tonomura86}. 
In other words, the shielding in the experiment of Ref.~\onlinecite{tonomura86}
was only one-sided where the incident electron is moving in a field-free
region, whereas shielding of both sides is necessary to eliminate the
Aharonov-Bohm effect. 
The experimental result of Ref.~\onlinecite{tonomura86}
can be fully understood in the framework of the local field interaction between
the localized flux and the magnetic field produced by an incident electron.

\section{Conclusion}
A local-interaction-based theory of the ABC effect has recently
been formulated~\cite{kang13}; this theory is consistent with 
all the results predicted in the standard potential-based framework. 
To verify the local nature of the ABC effect, we have investigated the
interaction of a charge and a fluxon when an ideal conducting barrier
is placed between the two objects. In the classical treatment of shielding,
the ABC effect vanishes in the absence of the overlap of electromagnetic
fields for any geometry of the system. In quantum treatments of superconducting
barriers, however, the result depends on the geometry of the system. 
The superconducting shield suppresses the ABC interference completely 
in Configuration I~(Fig.~2(a)). In contrast, the ABC phase factor is
modified in Configuration II, or even unaffected in Configuration III. 
We have shown that the effect of shielding is determined
by the fluctuating local
interaction of the electromagnetic fields. Our study shows that 
the framework of local field
interaction is fully adequate for a universal description of 
the ABC effect. 

%
%
%
\bibliography{ref-ab}

\begin{thebibliography}{11}
\bibitem [{flo()}]{floating-sc}%
  \BibitemOpen
  \href@noop {} {}\bibinfo {note} {An exceptional case is Configuration II with
  a floating superconductor. In this case, the electric field of the charge is
  not compensated by the superconductor and extends to the location of the
  fluxon. Therefore, this is irrelevant to our analysis, and we consider only a
  grounded superconductor that perfectly shields the fields.}\BibitemShut
  {Stop}%
\end{thebibliography}

\begin{figure}[l]
\includegraphics[width=1.6in]{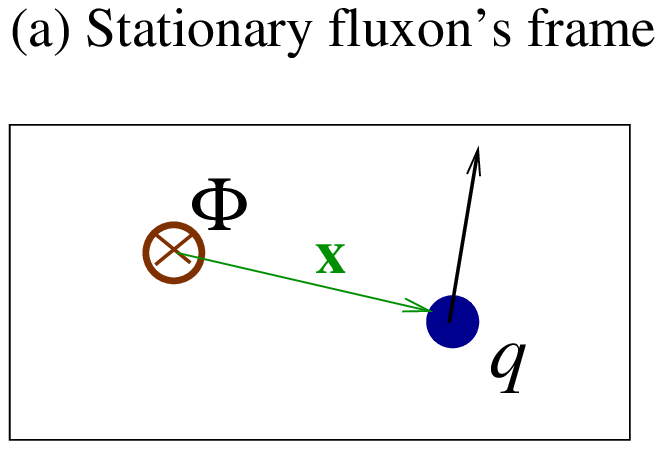}
\includegraphics[width=1.6in]{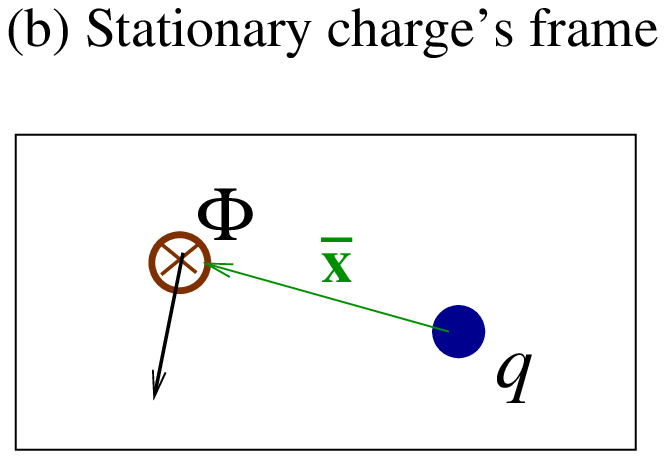} \\[0.3cm]
\includegraphics[width=1.6in]{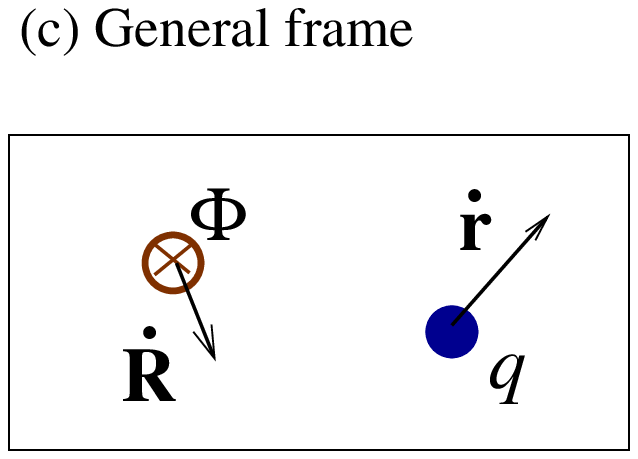}
\caption{A charge $q$ and a fluxon $\Phi$ moving in two spatial dimension in
 (a) the stationary fluxon's frame, (b) the stationary charge's frame, and
 (c) a general frame, respectively.
 }
\end{figure}

\begin{figure}
\includegraphics[width=1.5in]{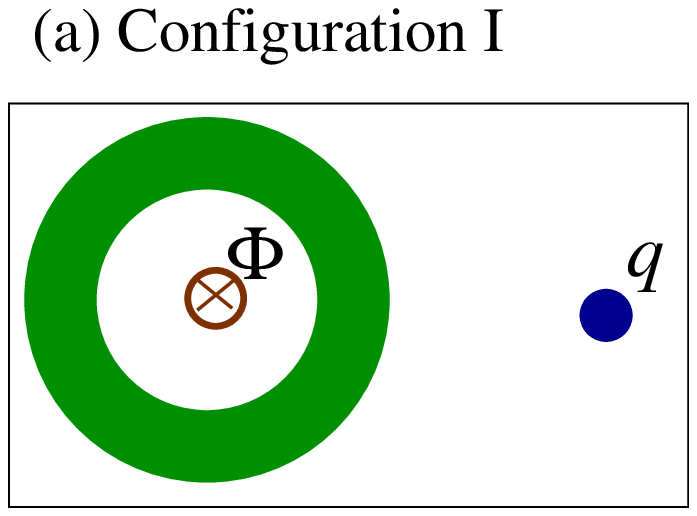} \hspace*{0.5cm}
\includegraphics[width=1.5in]{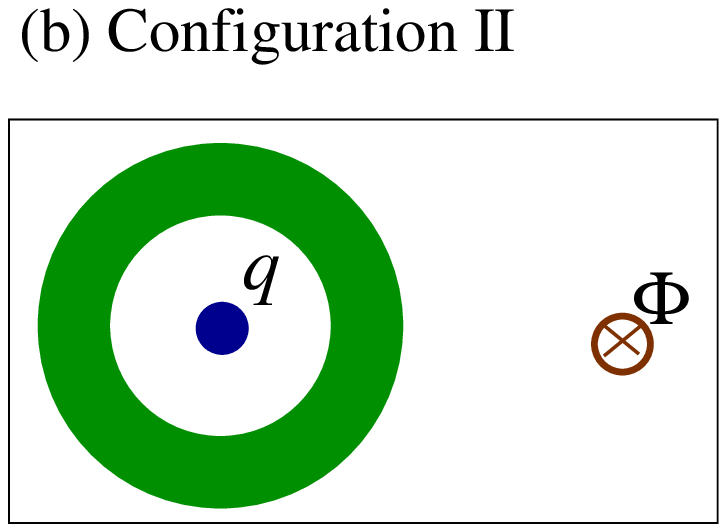} \\[0.3cm]
\includegraphics[width=1.5in]{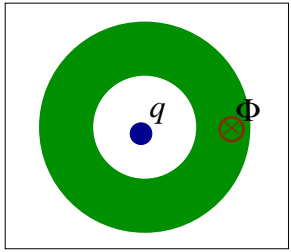}
\caption{Three possible configurations of shielding the overlap of 
 the electromagnetic fields produced by charge $q$ and fluxon $\Phi$.
 Green regions represent the superconducting barriers.
 }
\end{figure}
\begin{figure}
\includegraphics[width=2.8in]{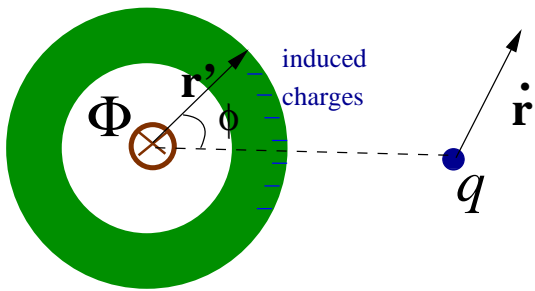}
\caption{The structure of the shielding in Configuration I of Fig.~2(a).
 }
\end{figure}

\end{document}